\documentclass{iopart}
\usepackage{graphicx}% Include figure files
\begin{document}
\title[{\it{\small Invited talk at the Workshop on Strongly
Correlated Electrons, Loughborough, U.K.}}]{Sharply increasing
effective mass: a precursor of a spontaneous spin polarization in a
dilute two-dimensional electron system}
\author{A~A Shashkin\dag\ddag, S~V Kravchenko\dag\footnote[7]{To whom
correspondence should be addressed}, V~T Dolgopolov\ddag\ and T~M
Klapwijk$\|$ }
\address{\dag\ Physics Department, Northeastern University, Boston,
MA 02115, U.S.A.}
\address{\ddag\ Institute of Solid State Physics, Chernogolovka,
Moscow District 142432, Russia}
\address{$\|$ Department of Applied Physics, Delft University of
Technology, 2628 CJ Delft, The Netherlands}
\begin{abstract}
We have measured the effective mass, $m$, and Land\'e $g$ factor in very dilute two-dimensional electron systems in silicon. Two independent methods have been used: (i)~measurements of the magnetic field required to fully polarize the electrons' spins and (ii)~analysis of the Shubnikov-de~Haas oscillations. We have observed a {\em sharp increase of the effective mass} with decreasing electron density while the $g$ factor remains nearly constant and close to its value in bulk silicon. The corresponding strong rise of the spin susceptibility $\chi\propto gm$ may be a precursor of a spontaneous spin polarization; unlike in the Stoner scenario, it originates from the enhancement of the effective mass rather than from the increase of $g$ factor. Furthermore, using tilted magnetic fields, we have found that the enhanced effective mass is independent of the degree of spin polarization and, therefore, its increase is not related to spin exchange effects, in contradiction with existing theories.  Our results show that the dilute 2D electron system in silicon behaves well beyond a weakly interacting Fermi liquid.
\end{abstract}


\ead{s.kravchenko@neu.edu}
%\maketitle

\section{Introduction}

At sufficiently low electron densities, two-dimensional (2D) electron
systems become strongly correlated, because the kinetic energy is
overpowered by energy of electron-electron interactions. The strength
of the interactions is usually characterized by the ratio between the
Coulomb energy and the Fermi energy, $r_s=E_c/E_F$, which, assuming
that the effective electron mass is equal to the band mass, in the
systems with single-valley spectrum reduces to the Wigner-Seitz
radius, $1/(\pi n_s)^{1/2}a_B$ (here $n_s$ is the electron density
and $a_B$ is the Bohr radius in semiconductor). There are several
suggested candidates for the ground state of the system, for example,
(i)~Wigner crystal characterized by spatial and spin ordering
\cite{wigner34}, (ii)~ferromagnetic Fermi liquid with spontaneous
spin ordering \cite{stoner47}, and (iii)~paramagnetic Fermi liquid
\cite{landau57}. In the strongly-interacting limit ($r_s\gg1$), no
analytical theory has been developed to date. According to numeric
simulations \cite{tanatar89}, Wigner crystallization is expected in a
very dilute regime, when $r_s$ reaches approximately 35. The refined
numeric simulations \cite{attaccalite02} have predicted that prior to
the crystallization, in the range of the interaction parameter
$25<r_s<35$, the ground state of the system is a strongly correlated
ferromagnetic Fermi liquid. A paramagnetic Fermi liquid is realized
at yet higher electron densities when the interactions are relatively
weak ($r_s\sim1$). The effective mass, $m$, and Land\'e $g$ factor
within the Fermi liquid theory are renormalized due to spin exchange
effects, with renormalization of the $g$ factor being dominant
compared to that of the effective mass \cite{renorm}. Alternatively, near the onset of Wigner crystallization, strong increase of the effective mass is expected \cite{spivak01,dolgopolov02}.

Recently, there has been a lot of interest in the electron properties
of dilute 2D systems due to the observation of an unexpected
metal-insulator transition (MIT) in zero magnetic field, strong
metallic temperature dependence of the resistance in these systems,
and a giant positive magnetoresistance in a magnetic field parallel
to the 2D plane (for a review, see Ref.~\cite{abrahams01}).  The most
pronounced effects have been observed in high-mobility silicon
metal-oxide-semiconductor field-effect transistors (MOSFETs), with
the low-temperature drop of the resistance reaching an order of
magnitude and magnetoresistance exceeding five orders of magnitude.
Significant progress has been recently made in understanding the
metallic behaviour of the  resistivity and its suppression by a
magnetic field \cite{zala01,punnoose02}; the metal-insulator
transition, however, still lacks adequate theoretical description.

In this paper, which summarizes results obtained in
Refs.~\cite{kravchenko00,shashkin01,shashkin02,comment,shashkin03},
we report measurements of the effective mass, $m$, and Land\'e $g$
factor in a wide range of electron densities including the immediate
vicinity of the MIT.  We have used two independent methods to
determine these parameters.  In the first method, we have studied
low-temperature magnetotransport in a parallel magnetic field.  It
turns out that the magnetic field, $B_c$, required to fully polarize
electron spins, is a strictly linear function of the electron density:
$B_c\propto(n_s-n_\chi)$ where $n_\chi$ is some finite electron
density close to the critical electron density $n_c$ for the $B=0$
metal-insulator transition.  Vanishing $B_c$ points to a sharply
increasing spin susceptibility, $\chi\propto gm$, and gives evidence
in favour of the spontaneous spin polarization at a finite electron
density.  (Similar conclusion about possible spontaneous spin
polarization in the dilute 2D electron system in silicon has been reached in
Ref.~\cite{vitkalov01}.)  Comparing our data for zero-field
resistivity with the recent theory \cite{zala01}, we extract the
values of $m$ and $g$ separately.  It turns out that it is the value
of the effective mass that becomes strongly (by more than a factor of
3) enhanced with decreasing electron density, while the $g$ factor
remains nearly constant and close to its value in bulk silicon.  In
the second method, we have determined the effective mass by analyzing
temperature dependence of the weak-field Shubnikov-de~Haas (SdH)
oscillations and found good agreement with the data obtained by the
first method.  Furthermore, using tilted magnetic fields, we find
that the value of the effective mass does not depend on the degree of
spin polarization, which points to a {\em spin-independent} origin of
the effective mass enhancement. This is in clear contradiction with
existing theories \cite{renorm,spivak01,dolgopolov02}.
\section{Samples and experimental technique}

Measurements have been made in a rotator-equipped Oxford dilution
refrigerator with a base temperature of $\approx30$~mK on
low-disordered (100)-silicon samples similar to those previously used
in Ref.~\cite{heemskerk98}.  Peak electron mobilities in these
samples are close to 3~m$^2$/Vs at 0.1~K.  To minimize the contact
resistance, which tends to grow very high at mK temperatures and low
electron density, thin gaps in the gate metallization have been
introduced; this allows for maintaining high electron density near
the contacts (about $1.5\times 10^{12}$~cm$^{-2}$) regardless of its
value in the main part of the sample.  These gaps are narrow enough
($<100$~nm) for the given gate-oxide thickness to provide a smoothly
descending electrostatic potential from the high-density part to the
low-density part.  The resistance, $R_{xx}$, has been measured by a
standard 4-terminal technique at a low frequency (0.4~Hz) to minimize
the out-of-phase signal.  Excitation current has been kept low
(0.1-0.2~nA) to ensure that measurements are taken in the linear
regime of response; the power generated in the samples has been
maintained under $10^{-14}$~W.  To verify that the electrons are not
overheated, we have studied the temperature dependence of the
amplitude of the SdH oscillations; the latter has been
found to follow the theoretical curve down to temperatures less than
50~mK (for more on this, see below).

\begin{figure}\vspace{-25.2mm}\hspace{-3mm}
\scalebox{.7}{\includegraphics{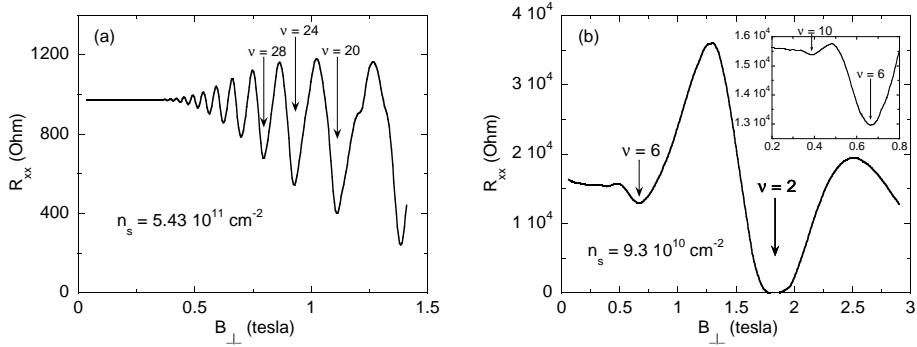}}\vspace{-119mm}
\caption{\label{1} Shubnikov-de~Haas oscillations in the Si MOSFET at
$T\approx 40$~mK (a) at a relatively high electron density
$n_s=5.43\times 10^{11}$~cm$^{-2}$ and (b) at low electron density
$n_s=9.3\times 10^{10}$~cm$^{-2}$.  The minima of the resistance at
Landau level filling factors $\nu=$~6 and 10 are shown on an expanded
scale in the inset.}
\end{figure}

\section{Experimental results}

We start by showing a low-temperature longitudinal magnetoresistance
$R_{xx}$ in a perpendicular magnetic field $B_\perp$ for a relatively
high (Fig.~\ref{1}(a)) and relatively low (Fig.~\ref{1}(b)) electron
densities.  At the high density, positions of SdH oscillations
correspond to ``cyclotron'' filling factors \cite{spin}, some of which are marked by arrows.  Indeed, the energy splittings $\Delta_s=g\mu_BB_\perp$ at ``spin'' filling factors, $\nu=$~2, 6, 10...~$=4i-2$, in high-density Si MOSFETs are known to be much smaller than the splittings $\Delta_c=\hbar\Omega_c-g\mu_BB_\perp$ at ``cyclotron'' filling factors, $\nu=$~4, 8, 12...~$=4i$, disregarding the odd $\nu$ valley splitting which is small (here $\Omega_c=eB_\perp/mc$ is the cyclotron frequency and $i=$~1, 2, 3...).  The behaviour of the sample at a relatively high electron density is thus rather ordinary.  In contrast, at low electron density (just above the metal-insulator transition which in this sample occurs at $n_s=n_c=8\times10^{10}$~cm$^{-2}$), the magnetoresistance looks quite different \cite{diorio90}.  The resistance minima are seen only at $\nu=$~2, 6, and 10 (see the inset); there is also a minimum at $\nu=1$ (not shown in the figure) corresponding to the valley splitting.  There are neither dips nor other anomalies at magnetic fields corresponding to $\nu=$~4, 8, or 12 where cyclotron minima are expected.

\begin{figure}\vspace{-28.5mm}\hspace{-16mm}
\scalebox{0.8}{\includegraphics{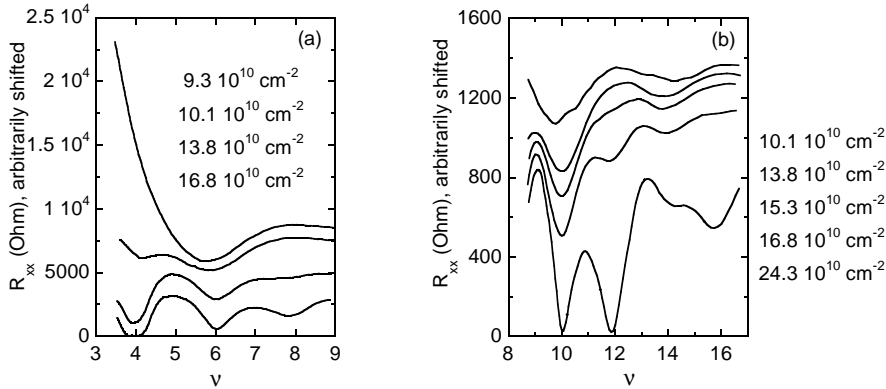}}\vspace{-14cm}
\caption{\label{2} Evolution of the Shubnikov-de~Haas oscillations
with electron density in two ranges of filling factors: (a) $3<\nu<9$
and (b) $8<\nu<17$; $T\approx 40$~mK. The curves are vertically
shifted for clarity.}
\end{figure}

Figure~\ref{2} shows how the resistance minima corresponding to the
cyclotron splittings gradually disappear as the electron density is
reduced.  At the highest electron densities (the lower curves), deep
resistance minima near even filling factors are seen ($\nu=$~4, 6,
and 8 in Fig.~\ref{2}(a); $\nu=$~10, 12, and 16 in Fig.~\ref{2}(b)),
and a shallow minimum is visible at $\nu=14$ in Fig.~\ref{2}(b).  As
$n_s$ is reduced, the minima at $\nu=$~4, 8, 12, and 16 become less
deep, and at the lowest electron densities (the upper curves),
neither of them is seen any longer, and only minima at $\nu=$~6, 10,
and 14 remain.

Our results thus show that as one approaches the metal-insulator
transition, the energy gaps at ``cyclotron'' filling factors become
gradually smaller than those at ``spin'' filling factors and
eventually vanish.  The condition for vanishing $\Delta_c$ is
$g\mu_BB_\perp=\hbar\Omega_c$ (within the uncertainty associated with the
broadening of the energy levels), or $gm/2m_e=1$, which is higher by
more than a factor of five than the ``normal'' value of this ratio,
$gm/2m_e=0.19$.  Therefore, the spin susceptibility $\chi\propto gm$
is strongly enhanced near the MIT.

One could attempt to link the observed behaviour to a many-body
enhancement of spin gaps specific for a perpendicular magnetic field
\cite{kallin84}.  However, the disappearance of the cyclotron
splittings in a wide range of magnetic fields would require an
enhanced $g$-factor which is independent of magnetic field, in
contradiction with Ref.~\cite{kallin84}.  On the other hand, our
results are consistent with the suggestion \cite{okamoto99} that the
effective $g$ factor is nearly field-independent and approximately
equal to its many-body enhanced zero-field value.  To probe this
conjecture, we have studied the {\em parallel-field} magnetotransport
in a wide range of electron densities.

\begin{figure}\vspace{-33mm}\hspace{23mm}
\scalebox{0.4}{\includegraphics{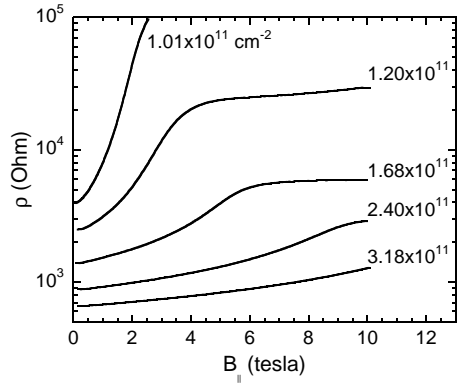}}\vspace{-2.5cm}
\caption{\label{3} Low-temperature magnetoresistance in parallel
magnetic fields at different electron densities above the critical
density for the $B=0$ metal-insulator transition. The lowest density
curve is outside the scaling region as described in the text.}
\end{figure}

Typical curves of the low-temperature magnetoresistance
$\rho(B_\parallel)$ in a parallel magnetic field are displayed in
Fig.~\ref{3}. Note that the thickness of the 2D electron system in Si
MOSFETs is small compared to the magnetic length in accessible
fields, and, therefore, the parallel field couples largely to the
electrons' spins while the orbital effects are suppressed
\cite{simonian97,rem1}.  The resistivity increases with field until
it saturates at a constant value above a certain density-dependent
magnetic field. According to Refs.~\cite{okamoto99,vitkalov00}, the
saturation of the magnetoresistance indicates the onset of a complete
spin polarization.  In the vicinity of the metal-insulator
transition, the magnetoresistance is strongly $T$-dependent down to
the lowest achievable temperatures. As one moves away from the
transition, however, the temperature dependences saturate at very low
temperatures. The data shown below are obtained in this
low-temperature limit where the magnetoresistance becomes
temperature-independent.

\begin{figure}\vspace{-35.0mm}\hspace{23mm}
\scalebox{0.4}{\includegraphics{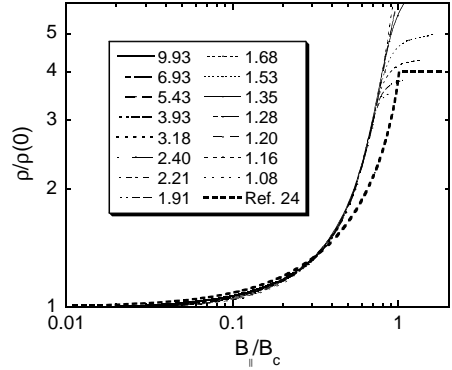}}\vspace{-2.5cm}
\caption{\label{4} Scaled curves of the normalized magnetoresistance
at different $n_s$ vs.\ $B_\parallel/B_c$. The electron densities are
indicated in units of $10^{11}$~cm$^{-2}$. Also shown by a dashed
line is the normalized magnetoresistance calculated in
Ref.~\protect\cite{dolgopolov00}.}
\end{figure}

In Fig.~\ref{4}, we show how the normalized magnetoresistance,
measured at different electron densities, collapses onto a single
curve when plotted as a function of $B_\parallel/B_c$. The scaling
parameter, $B_c$, has been normalized to correspond to the magnetic
field at which the magnetoresistance saturates (within the accuracy
with which the latter can be determined). The observed scaling is
remarkably good for $B_\parallel/B_c\le 0.7$ in the electron density
range $1.08\times10^{11}-10^{12}$~cm$^{-2}$, although with increasing
$n_s$, the scaled experimental data occupy progressively shorter
intervals on the resulting curve. Both at $B_\parallel/B_c>0.7$ and
outside the indicated range of electron densities, the scaled data
start to noticeably deviate from the universal curve. In particular,
the scaling breaks down when one approaches ($n_s<1.3\,n_c$) the
metal-insulator transition which in this sample occurs at zero
magnetic field at $n_c=8\times 10^{10}$~cm$^{-2}$. This is not
surprising as the magnetoresistance near $n_c$ depends strongly on
temperature, as discussed above. We note that the observed scaling
dependence is described reasonably well by the theoretical dependence
of $\rho/\rho(0)$ on the degree of spin polarization
$\xi=gm\mu_BB_\parallel/\pi\hbar^2n_s=B_\parallel/B_c$ predicted by
the recent theory \cite{dolgopolov00}.

\begin{figure}\vspace{-34mm}\hspace{23mm}
\scalebox{0.4}{\includegraphics{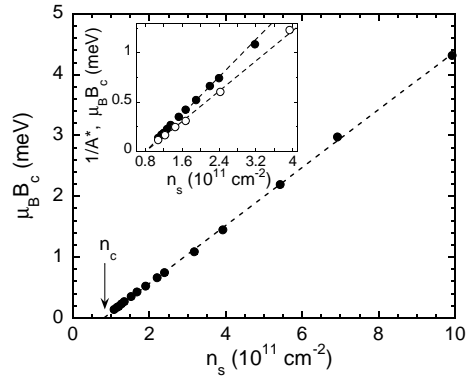}}\vspace{-2.5cm}
\caption{\label{5} Dependence of the field $B_c$ on electron density.
The dashed line is a linear fit which extrapolates to the critical
electron density for the $B=0$ metal-insulator transition.
Comparison of $1/A^*(n_s)$ (see Eq.~(\ref{sigma})) and $B_c(n_s)$ (open
and solid circles, correspondingly) is shown in the inset.  The
dashed lines are linear fits which extrapolate to the critical
electron density for the metal-insulator transition.}
\end{figure}

In Fig.~\ref{5}, $B_c$ is plotted vs.\ $n_s$.  With high accuracy,
$B_c$ is proportional to the deviation of the electron density from
its critical value, {\it i.e.}, to $(n_s-n_c)$, over a wide range of
electron densities. In other words, the field, at which the
magnetoresistance saturates, tends to vanish at $n_c$ (see also
Refs.~\cite{comment,vitkalov02}). We emphasize that our procedure
provides high accuracy for determining the behaviour of the field of
saturation with electron density, {\it i.e.}, the functional form of
$B_c(n_s)$, even though the absolute value of $B_c$ is determined not
so accurately. Note that at $n_s$ above $2.4\times
10^{11}$~cm$^{-2}$, the saturation of the resistance is not reached
in our magnetic field range; still, the high precision of the
collapse of the high-density experimental curves onto the same
scaling curve as the low-density data allows us to draw conclusions
about the validity of the obtained law $B_c(n_s)$ over a much wider
range of electron densities.

The observed tendency of $B_c$ to vanish at a finite electron density
is consistent with the strong increase of the spin susceptibility
$\chi\propto gm$ \cite{kravchenko00,comment,pudalov02} and gives
evidence in favour of a spontaneous spin polarization at
$n_\chi\approx n_c$. In principle, either $g$ or $m$ (or both) may be
responsible for the strong increase of the spin susceptibility.  As
has already been mentioned, within the Fermi liquid theory, both the
effective mass and $g$ factor are renormalized due to spin exchange
effects, with renormalization of the $g$ factor being dominant
compared to that of the effective mass.  In contrast, the dominant
increase of the effective mass follows from an alternative
description of the strongly-interacting electron system beyond the
Fermi liquid approach \cite{spivak01,dolgopolov02}.  To separate $g$
and $m$, we have measured the temperature-dependent conductivity in
zero magnetic field and analyzed the data in the spirit of recent
theory \cite{zala01}.  According to this theory, $\sigma$ is a linear
function of temperature:

\begin{equation}
\frac{\sigma(T)}{\sigma_0}=1-A^*k_BT, \label{sigma}\end{equation}
where the slope, $A^*$, is determined by the interaction-related
parameters: the Fermi liquid constants, $F_0^a$ and $F_1^s$:

\begin{equation}
A^*=-\frac{(1+\alpha F_0^a)gm}{\pi\hbar^2n_s}.
\label{A^*}\end{equation} The factor $\alpha$ is equal to 8 in our
case \cite{private}. This theoretical relation allows us to determine
the many-body enhanced $g$ factor and mass $m$ separately using the
data for the slope $A^*$ and the product $gm$.

\begin{figure}\vspace{-34.3mm}\hspace{28mm}
\scalebox{0.4}{\includegraphics{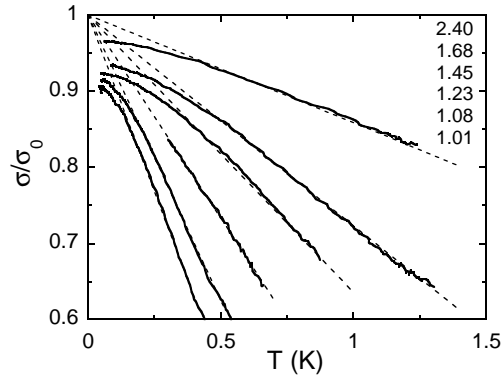}}\vspace{-2.5cm}
\caption{\label{6} The temperature dependence of the normalized
conductivity at different electron densities (indicated in units of
$10^{11}$cm$^{-2}$) above the critical electron density for the
metal-insulator transition. The dashed lines are fits of the linear
interval of the dependence.}
\end{figure}

Typical dependences of the normalized conductivity on temperature,
$\sigma(T)/\sigma_0$, are displayed in Fig.~\ref{6} at different
electron densities above $n_c$; the value $\sigma_0$, which has been
used to normalize $\sigma$, was obtained by extrapolating the linear
interval of the $\sigma(T)$ dependence to $T=0$. As long as the
deviation $|\sigma/\sigma_0-1|$ is sufficiently small, the
conductivity $\sigma$ increases linearly with decreasing $T$ in
agreement with Eq.~(\ref{sigma}), until it saturates at the lowest
temperatures. As seen from the figure, the linear interval of the
dependence is wide enough to make a reliable fit.

The $n_s$ dependence of the inverse slope $1/A^*$, extracted from the
$\sigma(T)$ data, is shown in the inset to Fig.~\ref{5} by open
circles.  Over a wide range of electron densities, the values $1/A^*$
and $\mu_BB_c$ turn out to be close to each other. The low-density
data for $1/A^*$ are approximated well by a linear dependence which
extrapolates to the critical electron density $n_c$ in a similar way
to the behaviour of the polarization field $B_c$.

\begin{figure}\vspace{-9.4mm}\hspace{23mm}
\scalebox{0.4}{\includegraphics{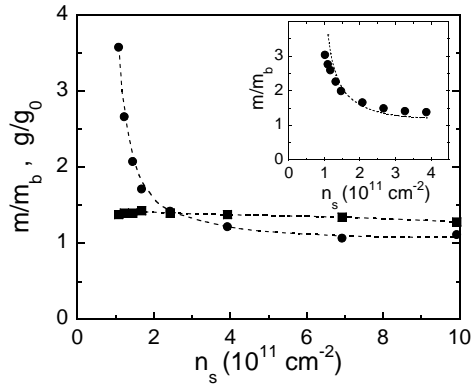}}\vspace{-4.8cm}
\caption{\label{7} The effective mass (circles) and $g$ factor
(squares), determined from the analysis of the parallel field
magnetoresistance and temperature-dependent conductivity, vs.\ electron
density.  The dashed lines are guides to the eye.  The inset compares
the so-obtained effective mass (dotted line) with the one extracted
from the analysis of SdH oscillations (circles).}
\end{figure}

In Fig.~\ref{7}, we show the so-determined values $g/g_0$ and $m/m_b$
as a function of the electron density (here $g_0=2$ is the $g$ factor
in bulk silicon, $m_b$ is the band mass equal to $0.19m_e$, and $m_e$ 
is the free electron mass).  In the high $n_s$ region
(relatively weak interactions), the enhancement of both $g$ and $m$
is relatively small, both values slightly increasing with decreasing
electron density in agreement with earlier data \cite{ando82}. Also,
the renormalization of the $g$ factor is dominant compared to that of
the effective mass, which is consistent with theoretical studies
\cite{renorm}.

In contrast, the renormalization at low $n_s$ (near the critical
region), where $r_s\gg1$, is much more striking. As the electron
density is decreased, the renormalization of the effective mass
overshoots abruptly while that of the $g$ factor remains relatively
small, $g\approx g_0$, without tending to increase. Hence, the
current analysis indicates that it is the effective mass, rather than
the $g$ factor, that is responsible for the drastically enhanced $gm$
value near the metal-insulator transition.

\begin{figure}\vspace{-0.88in}\hspace{26mm}
\scalebox{0.38}{\includegraphics{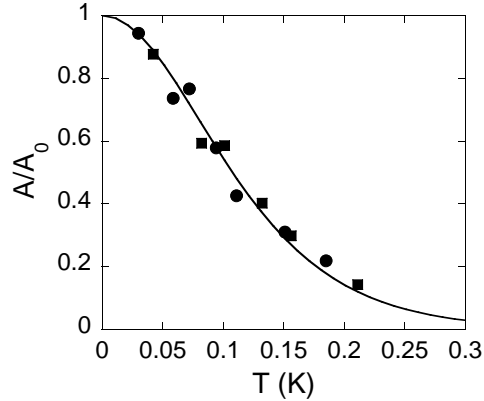}}\vspace{-1.1in}
\caption{\label{8} Amplitude of the weak-field SdH oscillations vs.\
temperature at $n_s=1.17\times 10^{11}$~cm$^{-2}$ for oscillation
numbers $\nu=hcn_s/eB_\perp=10$ (dots) and $\nu=14$ (squares). The
value of $T$ for the $\nu=10$ data is divided by the factor of 1.4.
The solid line is a fit using Eq.~(\ref{A}).}
\end{figure}

Since the procedure for extracting $g$ and $m$ described above relies
on theoretically calculated functional form for the slope $A^*$
\cite{zala01}, we have performed independent measurements of the
effective mass based on the temperature analysis of the amplitude,
$A$, of the weak-field (sinusoidal) SdH oscillations.  A typical
temperature dependence of $A$ for the normalized resistance,
$R_{xx}/R_0$ (where $R_0$ is the average resistance), is displayed in
Fig.~\ref{8}. To determine the effective mass, we use the method of
Ref.~\cite{smith72} extending it to much lower electron densities and
temperatures. We fit the data for $A(T)$ using the formula

\begin{equation}
A(T)=A_0\frac{2\pi^2k_BT/\hbar\Omega_c}{\sinh(2\pi^2k_BT/\hbar
\Omega_c)},\end{equation}\label{A}
where $A_0=4\exp(-2\pi^2k_BT_D/\hbar\Omega_c)$ and $T_D$ is
the Dingle temperature. As the latter is related to the level width
through the expression $T_D=\hbar/2\pi k_B\tau$ (where $\tau$ is the
elastic scattering time) \cite{ando82}, damping of the SdH
oscillations with temperature may be influenced by
temperature-dependent $\tau$.  We have verified that in the studied
low-temperature limit for electron densities down to $\approx 1\times
10^{11}$~cm$^{-2}$, possible corrections to the mass value caused by
the temperature dependence of $\tau$ (and hence $T_D$) are within our
experimental uncertainty which is estimated at about 10\%. Note that
the amplitude of the SdH oscillations follows the calculated curve
down to the lowest achieved temperatures, which confirms that the
electrons were in a good thermal contact with the bath and were not
overheated.  The fact that the experimental dependence $A(T)$ follows
the theoretical curve justify applicability of Eq.~(\ref{A}) to this
strongly-interacting electron system.

The so-determined effective mass is shown in the inset to
Fig.~\ref{7}.  In quantitative agreement with the results obtained by
the alternative method described above (the dotted line), the
effective mass sharply increases with decreasing $n_s$.  The
agreement between the results obtained by two independent methods
adds confidence in our results and conclusions.  Our data are also
consistent with the data for spin and cyclotron gaps obtained by
magnetocapacitance spectroscopy \cite{khrapai03}.

\begin{figure}\vspace{-0.9in}\hspace{26mm}
\scalebox{0.38}{\includegraphics{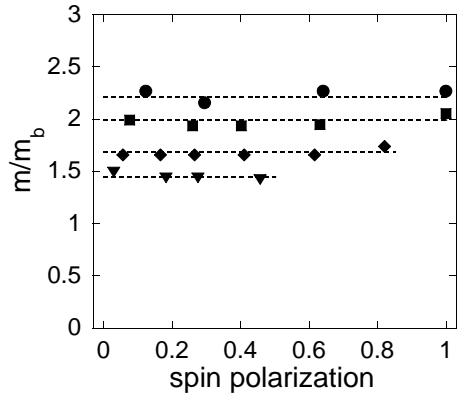}}\vspace{-1.1in}
\caption{\label{9} The effective mass vs.\ the degree of spin
polarization for the following electron densities in units of
$10^{11}$~cm$^{-2}$: 1.32 (dots), 1.47 (squares), 2.07 (diamonds),
and 2.67 (triangles). The dashed lines are guides to the eye.}
\end{figure}

A strong enhancement of $m$ at low electron densities may originate
from spin effects \cite{renorm,spivak01,dolgopolov02}. With the aim
of probing a possible contribution from the spin effects, we have
introduced a parallel magnetic field component to align the
electrons' spins.  In Fig.~\ref{9}, we show the behaviour of the
effective mass with the degree of spin polarization,
$p=(B_\perp^2+B_\parallel^2)^{1/2}/B_c$. As seen from the figure,
within our accuracy, the {\em effective mass $m$ does not depend on
$p$}. Therefore, the $m(n_s)$ dependence is robust, the origin of the
mass enhancement has no relation to the electrons' spins and exchange
effects \cite{rem2}.

\section{Discussion}

Under the conditions of our experiments, the interaction parameter,
$r_s$, is larger by a factor of $2m/m_b$ than the Wigner-Seitz radius
and reaches approximately 50, which is above the theoretical estimate
for the onset of Wigner crystallization. As has already been
mentioned, two approaches to calculate the renormalization of $m$ and
$g$ have been formulated. The first one exploits the Fermi liquid
model extending it to relatively large $r_s$. Its main outcome is
that the renormalization of $g$ is large compared to that of $m$
\cite{renorm}. In the limiting case of high $r_s$, one may expect a
divergence of the $g$ factor that corresponds to the Stoner
instability. These predictions are in obvious contradiction to our
data: (i) the behaviour of the 2D dilute system in the regime of the
strongly enhanced susceptibility --- close to the onset of
spontaneous spin polarization and Wigner crystallization --- is
governed by the effective mass, rather than the $g$ factor, through
the interaction parameter $r_s$; and (ii) the insensitivity of the
effective mass to spin effects also cannot be accounted for.

The other theoretical approach either employs analogy between a
strongly interacting 2D electron system and He$^3$ \cite{spivak01} or
applies Gutzwiller's variational method \cite{brinkman70} to Si
MOSFETs \cite{dolgopolov02}. It predicts that near the
crystallization point, the renormalization of $m$ is dominant
compared to that of $g$ and that the effective mass tends to diverge at the
transition. Although the sharp increase of the mass is in agreement
with our findings, it is the expected dependence of $m$ on the degree
of spin polarization that is not confirmed by our data: the model of
Ref.~\cite{spivak01} predicts that the effective mass should increase
with increasing spin polarization, whereas the prediction of the
other model \cite{dolgopolov02} is the opposite.

Thus, the existing theories fail to explain our finding that in a
dilute 2D electron system the effective mass is strongly enhanced and
does not depend on the degree of spin polarization. The fact that the
spin exchange is not responsible for the observed mass enhancement
reduces the chances for the occurrence of the ferromagnetic Fermi
liquid prior to the Wigner crystallization.  However, should the spin
exchange be small, the spin effects may still come into play closer
to the onset of Wigner crystallization where the Fermi energy may
continue dropping as caused by mass enhancement.

In summary, we have found that in very dilute two-dimensional
electron systems in silicon, the effective mass sharply increases
with decreasing electron density, while the $g$ factor remains nearly
constant and close to its value in bulk silicon.  The enhanced
effective mass does not depend on the degree of the spin polarization
and, therefore, its increase is not related to spin exchange effects,
in contradiction with existing theories.  The corresponding strong
rise of the spin susceptibility may be a precursor of a spontaneous
spin polarization; unlike in the Stoner scenario, the latter
originates from the enhancement of the effective mass rather than the
increase of the $g$ factor.  Our results show that the dilute 2D
electron system in silicon behaves well beyond the weakly interacting
Fermi liquid.

\section*{Acknowledgments}
We gratefully acknowledge discussions with I.~L. Aleiner, D. Heiman,
F. Kusmartsev, M.~P. Sarachik, B. Spivak, and S.~A. Vitkalov.  This
work was supported by the National Science Foundation grants
DMR-9988283 and DMR-0129652, the Sloan Foundation, the Russian
Foundation for Basic Research, the Russian Ministry of Sciences, and
the Programme ``The State Support of Leading Scientific Schools''.

%\newpage


\section*{References}
\begin{thebibliography}{10}
\bibitem{wigner34} E. Wigner, Phys.\ Rev.\ {\bf 46}, 1002 (1934).
\bibitem{stoner47} E.~C. Stoner, Rept.\ Prog.\ Phys.\ {\bf 11}, 43
(1947).
\bibitem{landau57} L.~D. Landau, Sov.\ Phys.\ JETP\ {\bf 3}, 920
(1957).
\bibitem{tanatar89} B. Tanatar and D.~M. Ceperley, Phys.\ Rev.\ B\
{\bf 39}, 5005 (1989).
\bibitem{attaccalite02} C. Attaccalite, S. Moroni, P. Gori-Giorgi,
and G.~B. Bachelet, Phys.\ Rev.\ Lett.\ {\bf 88}, 256601 (2002).
\bibitem{renorm} N. Iwamoto, Phys.\ Rev.\ B\ {\bf 43}, 2174 (1991);
Y. Kwon, D.~M. Ceperley, and R.~M. Martin, {\it ibid.} {\bf 50}, 1684
(1994); G.-H. Chen and M.~E. Raikh, {\it ibid.} {\bf 60}, 4826
(1999).
\bibitem{spivak01} B. Spivak, Phys.\ Rev.\ B\ {\bf 64}, 085317
(2001).
\bibitem{dolgopolov02} V.~T. Dolgopolov, JETP\ Lett.\ {\bf 76}, 377
(2002).
\bibitem{abrahams01} E. Abrahams, S.~V. Kravchenko, and M.~P.
Sarachik, Rev.\ Mod.\ Phys.\ {\bf 73}, 251 (2001).
\bibitem{zala01} G. Zala, B.~N. Narozhny, and I.~L. Aleiner, Phys.\
Rev.\ B\ {\bf 64}, 214204 (2001).
\bibitem{punnoose02} A. Punnoose and A.~M. Finkelstein, Phys.\ Rev.\
Lett.\ {\bf 88}, 016802 (2002).
\bibitem{kravchenko00} S.~V. Kravchenko, A.~A. Shashkin, D.~A.
Bloore, and T.~M. Klapwijk, Solid\ State\ Commun.\ {\bf 116}, 495
(2000).
\bibitem{shashkin01} A.~A. Shashkin, S.~V. Kravchenko, V.~T.
Dolgopolov, and T.~M. Klapwijk, Phys.\ Rev.\ Lett.\ {\bf 87}, 086801
(2001).
\bibitem{shashkin02} A.~A. Shashkin, S.~V. Kravchenko, V.~T.
Dolgopolov, and T.~M. Klapwijk, Phys.\ Rev.\ B\ {\bf 66}, 073303
(2002).
\bibitem{comment} S.~V. Kravchenko, A.~A. Shashkin, and V.~T.
Dolgopolov, Phys.\ Rev.\ Lett.\ {\bf 89}, 219701 (2002).
\bibitem{shashkin03} A.~A. Shashkin, M. Rahimi, S. Anissimova, S.~V.
Kravchenko, V.~T. Dolgopolov, and T.~M. Klapwijk, preprint
cond-mat/0301187.
\bibitem{vitkalov01} S.~A. Vitkalov, H. Zheng, K.~M. Mertes, M.~P.
Sarachik, and T.~M. Klapwijk, Phys.\ Rev.\ Lett.\ {\bf 87}, 086401
(2001).
\bibitem{heemskerk98} R. Heemskerk and T.~M. Klapwijk, Phys.\ Rev.\
B\ {\bf 58}, R1754 (1998).
\bibitem{spin} In silicon MOSFETs, ``cyclotron'' gaps correspond to $\nu=$~4, 8, 12, 16... while ``spin'' gaps correspond to $\nu=$~2, 6, 10, 14... due to a two-fold valley degeneracy in this system.
\bibitem{diorio90} This unusual behaviour was first reported by M. D'Iorio, V.~M. Pudalov, and S.~G. Semenchinsky, Phys.\ Lett.\ A\ {\bf 150}, 422 (1990).
\bibitem{kallin84} Yu.~A. Bychkov, S.~V. Iordanskii, and G.~M.
Eliashberg, JETP Lett.\ {\bf 33}, 143 (1981); C. Kallin and B.~I.
Halperin, Phys.\ Rev.\ B\ {\bf 30}, 5655 (1984); A.~H. MacDonald,
H.~C.~A. Oji, and K.~L. Liu, Phys.\ Rev.\ B {\bf 34}, 2681 (1986).
\bibitem{okamoto99} T. Okamoto, K. Hosoya, S. Kawaji, and A. Yagi,
Phys.\ Rev.\ Lett.\ {\bf 82}, 3875 (1999).
\bibitem{simonian97} D. Simonian, S.~V. Kravchenko, M.~P. Sarachik,
and V.~M. Pudalov, Phys.\ Rev.\ Lett.\ {\bf 79}, 2304 (1997).
\bibitem{rem1} The case for the 2D carrier system in GaAs is opposite
to that for Si, because the orbital effects in GaAs give rise to an
enhancement of the effective mass in parallel magnetic fields, see,
e.g., E. Tutuc {\it et al.}, cond-mat/0301027.
\bibitem{vitkalov00} S.~A. Vitkalov, H. Zheng, K.~M. Mertes, M.~P.
Sarachik, and T.~M. Klapwijk, Phys.\ Rev.\ Lett.\ {\bf 85}, 2164
(2000).
\bibitem{dolgopolov00} V.~T. Dolgopolov and A. Gold, JETP\ Lett.\
{\bf 71}, 27 (2000).
\bibitem{vitkalov02} S.~A. Vitkalov, M.~P. Sarachik, and T.~M.
Klapwijk, Phys.\ Rev.\ B\ {\bf 65}, 201106(R) (2002).
\bibitem{pudalov02} V.~M. Pudalov, M.~E. Gershenson, H. Kojima, N.
Butch, E.~M. Dizhur, G. Brunthaler, A. Prinz, and G. Bauer, Phys.\
Rev.\ Lett.\ {\bf 88}, 196404 (2002).
\bibitem{private} I.~L. Aleiner, private communication. For low
intervalley scattering, $\alpha=8$ if $T<\Delta_v$ and $\alpha=16$ if
$T\gg\Delta_v$, where $\Delta_v$ is the valley splitting. Both
experimental (see, e.g., V.~M. Pudalov, A. Punnoose, G. Brunthaler,
A. Prinz, and G. Bauer, cond-mat/0104347) and theoretical (see
Ref.~\cite{ando82}) studies give an estimate for $\Delta_v\approx
1.5$~K.
\bibitem{ando82} T. Ando, A.~B. Fowler, and F. Stern, Rev.\ Mod.\
Phys.\ {\bf 54}, 437 (1982).
\bibitem{smith72} J.~L. Smith and P.~J. Stiles, Phys.\ Rev.\ Lett.\
{\bf 29}, 102 (1972).
\bibitem{khrapai03} V.~S. Khrapai, A.~A. Shashkin, and V. T.
Dolgopolov, cond-mat/0301361.
\bibitem{rem2} In principle, the exchange effects can also originate
from the isospin degree of freedom in bivalley (100)-Si MOSFETs. The
valley origin of the strongly enhanced effective mass is not very
likely, as inferred from a similar increase of the ratio of the spin
and the cyclotron splittings at low $n_s$ in the 2D electron system
in GaAs \cite{zhu03}.
\bibitem{zhu03} J. Zhu, H.~L. Stormer, L.~N. Pfeiffer, K.~W. Baldwin,
and K.~W. West, Phys.\ Rev.\ Lett.\ (2003, in press).
\bibitem{brinkman70} W.~F. Brinkman and T.~M. Rice, Phys.\ Rev.\ B\
{\bf 2}, 4302 (1970).
\end{thebibliography}
\end{document}